# Deep ahead-of-threat virtual patching

Fady Copty[*], Andre Kassis, Sharon Keidar-Barner, Dov Murik

IBM Research – Haifa, Israel
[*]`fadyc@il.ibm.com`

**Abstract.** Many applications have security vulnerabilities that can be exploited. It is practically impossible to find all of them due to the NP-complete nature of the testing problem. Security solutions provide defenses against these attacks through continuous application testing, fast-patching of vulnerabilities, automatic deployment of patches, and virtual patching detection techniques deployed in network and endpoint security tools. These techniques are limited by the need to find vulnerabilities before the 'black hats'. We propose an innovative technique to virtually patch vulnerabilities before they are found. We leverage testing techniques for supervised-learning data generation, and show how artificial intelligence techniques can use this data to create predictive deep neural-network models that read an application's input and predict in real time whether it is a potential malicious input. We set up an ahead-of-threat experiment in which we generated data on old versions of an application, and then evaluated the predictive model accuracy on vulnerabilities found years later. Our experiments show ahead-of-threat detection on LibXML2 and LibTIFF vulnerabilities with 91.3% and 93.7% accuracy, respectively. We expect to continue work on this field of research and provide ahead-of-threat virtual patching for more libraries. Success in this research can change the current state of endless racing after application vulnerabilities and put the defenders one step ahead of the attackers.

**Keywords:** Virtual patching, Application vulnerability, Deep Learning.

## 1 Introduction

Every application has bugs and vulnerabilities. Some of them can be exploited by hackers, also known as 'black hats,' to gain control over the software, get credential information, or leak data. Both hackers and software developers invest a lot of effort in discovering these bugs. Hackers look for the next loophole, while developers try to detect and fix vulnerabilities before they are exploited. When hackers detect the vulnerability or when the application is not properly updated before the vulnerability is made public, we witness catastrophes such as the WannaCry campaign or Heartbleed bug.

Software verification and penetration testing are the usual means to detect these vulnerabilities. As software becomes more complex, verification has become an almost impossible challenge. Software testing methods cannot cover all possible scenarios, even with highly effective heuristics such as fuzz testing using genetic algorithms [2].

Software verification methods based on formal verification can cover all corner cases; however, they cannot scale to verify industrial-scale software.

How do hackers succeed despite the software testing applied? The "bad guys" have an advantage: once a vulnerable software is deployed it is not always patched immediately, even once a vulnerability is detected. This occurs because it takes time to design and implement a fix that will not affect the application's performance and user experience. The vulnerability may be in legacy code, which is hard to maintain, or the software may have been written by a third-party organization. Often the patch is simply delayed. According to Infosecurity Magazine [7], companies take an average of 100 to 120 days to patch vulnerabilities.

In the most difficult cases, the software cannot be patched because it is highly critical and cannot suffer from instability that may be associated with the fix. One example of this would be flight control software. Moreover, the software may be installed on a device that cannot be easily patched, such as a pacemaker.

The Internet of Things (IoT) is even harder to protect. When a patch is released, it must be deployed on many endpoints. Moreover, the many sensors and devices have an endless variation of software that needs to be installed, managed, and protected.

Solutions for virtual patches are introduced in cases where the software is not patched. Instead of patching the software, a virtual patch blocks malicious inputs that may exploit the vulnerabilities in the software. The virtual patch can be deployed as part of the firewall, or in the case of IoT, as part of the gateway. In network firewalls and proxies, such as ModSecurity and Snort, virtual patches are manually written rules that protect the software [13][0]. There is a configurable layer to which rules can be added. This capability makes the virtual patching process much easier than the traditional patching process, especially for cases in which the software cannot be patched at all. The main disadvantages of this method are that it requires human effort and can be applied only for known vulnerabilities.

Since software will never be free of bugs, mitigation systems such as Intrusion Detection Systems (IDS) [12] were developed to detect malicious software-input in real time. There are two types of IDS: misuse and anomaly detection systems. Misuse systems are based on signatures to avoid time consuming content inspection. For each malicious software-input, a signature is created and added to a black list. Any software-input with a signature on the blacklist is blocked. The problem is that IDS are not effective against variations of the malicious software-input; these can be easily created by making minor changes to the software-input or using different testing techniques. The effectiveness of an IDS is highly dependent on its ability to update signatures, which requires human effort. Anomaly-based IDS learns what constitutes normal behavior for the system and alerts when it detects a deviation from the normal behavior. Unfortunately, some types of attack appear as normal behavior. Both IDS methods have a low probability of detecting zero-day attacks. In addition, IDS suffer from a high rate of false positive and false negatives, which may lead them to block legitimate software or grant access to malicious software.

We introduce a novel, fully-automated, approach that provides ahead-of-threat virtual patching for security loopholes. This approach leverages fuzz testing technology, based on genetic algorithms, to create data for a machine learning algorithm. The algorithm then learns which inputs need to be blocked. No human effort is required to analyze the software vulnerabilities, no effort is required to label the training data, and prior knowledge on the vulnerabilities in the software is not mandatory. The virtual patch is a deep neural network (DNN) predictive model. It can be produced by an application developer from source code and used as an application hardening method. It can also be produced by a security vendor from source code or binary and deployed in a deep inspection IDS or endpoint protection tool for file scanning.

## 2  Related work

The fields of virtual patching and intrusion detection are heavily researched, both in the industry and in academia. Mishra et al. [12] provide an extensive survey of intrusion detection techniques applied to many levels in a cloud environment. According to their categorization, our approach is classified as a misuse detection technique with a decision engine trained by machine learning. This is as opposed to signature-based misuse detection, which is prone to overlooking novel attack patterns.

Many recent IDS implementations use machine learning techniques. Kim et al. [8] use SVM to build a machine-learning based IDS. Li et al. [9] propose an artificial neural network (ANN) based IDS. Pandeeswari et al. [14] employ a hybrid algorithm of Fuzzy C-Means clustering and ANN. Ashfaq et al. [3] use a semi-supervised learning approach to intrusion detection to reduce the number of labeled examples needed to train the model. Aljawarneh et al. [1] use anomaly detection and a hybrid approach of seven different ML algorithms to improve accuracy and maintain a low false positive rate. These authors base their decision on network traffic metadata. In contrast, our solution is trained on the actual data content (packet payload, file content) and can detect malformed, potentially malicious, packets and files. It can be used in addition to an IDS deployed at the network level or at the host level.

Our ability to detect malicious payloads using deep neural networks resembles work in the field of static malware detection using deep learning [15, 17]. However, most of these works rely on the availability of a large training set of benign and malicious samples, which are used to train the machine learning model. Our solution generates its own training set from a small corpus of examples and monitors the target application's execution on each sample input to decide that input's label.

## 3  Overview

Our solution takes an application's source code or binary, and automatically produces an ahead-of-threat virtual patch of the application. The virtual patch predicts–in real time—whether an input to an application will allow a vulnerability to be exploited in this application. The system can be divided into two components: (1) data set generation using testing techniques and (2) supervised learning using deep neural networks. The data set generation is described in section 3.1, and the supervised learning is described in sections 3.2 and 3.3. Sections 3.4 and 3.5 describe the training and evaluation methods.

### 3.1 Data set generation using testing techniques

Using deep learning techniques necessitates a large data set with diverse samples. Our work requires a large corpus of relevant inputs to the application that we are trying to protect. The corpus should include benign samples that the application can process, error samples that the application recognizes and rejects, and malicious samples that expose an application vulnerability. To build such a corpus, we used the software testing technique of fuzzing, which is usually used to find bugs or security vulnerabilities. We use a side effect of fuzzing: the millions of inputs that are generated during the fuzzing process. As part of our solution, we have the 'fuzzer' label these inputs as benign, error, or malicious, thereby forming a training set we can use to build our model.

Our data set generation effort relies on AFL [2] to generate the test cases and to exercise the application (running it with the generated inputs). We use AFL along with an address sanitizer to detect malicious tests. A test is labeled malicious if AFL shows that it crashes the application, or if it fails one of the sanitizer checkers. We made two modifications in AFL to support our requirements: we save non-unique test cases and save error cases as their own class. These modifications are described in detail below.

**Saving non-unique test cases.** AFL saves only unique input files; therefore, input files that hit an execution path that was already covered by other inputs are not added. While this is beneficial for creating a minimal corpus that covers the application's execution paths, we want to save many different inputs for each path to train our model. We added an option for AFL to save non-unique inputs.

**Saving error cases.** AFL categorizes inputs into two classes: benign and crash. Crashes cause abnormal program termination, often via an OS signal that indicates some memory access error. We added an option to detect error files by either recording the program exit code (where a non-zero exit code means the input is detected as an error) or by detecting output to the standard error stream (where output in standard error means the input is detected as invalid).

### 3.2 Automatic feature extraction

Typical IDS systems must operate in a real-time environment. This is the reason we selected feature-extraction methods that limit the inspection time to milliseconds, from the moment the file is received until a decision is made.

We deploy two techniques for feature extraction; counting the number of occurrences of a specific token in a file, and transforming the file byte sequence into a one-dimensional numerical vector and applying zero-padding. We explored the use of token bi-gram, but experiments showed that this exceeded our time limitations.

To create the token count features, we first determined what tokens to count. These tokens can be divided into two groups. The first group of tokens are any dictionary tokens received as optional input from the user. This group covers most of the expected input-language tokens. However, security vulnerabilities are often caused by strings not in this token group.

The second group of tokens is generated automatically using the genetic algorithm. A genetic test generation algorithm performs either a mutation or a crossover [18] on a chosen test to create a new test. We captured all successful mutations and used them as tokens. This type of token often corresponds to tokens that were not specified in the input-language specification, but were implemented in the application under test. This can happen due to poor specification or poor implementation.

### 3.3 Deep Learning

The structure of the deep neural network as depicted in **Fig. 1** is composed of two paths. These paths eventually combine into one output that predicts the probability of malicious versus benign. We used Keras Deep Learning Models with Scikit-Learn in Python [0][0] to implement this structure. The total number of parameters in our DNN model is around five million.

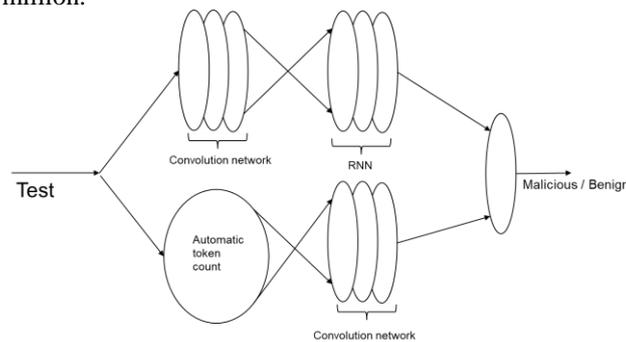

**Fig. 1.** Structure of deep neural network

In the first path of our deep neural network, we apply zero padding and transform every test into a numerical one-dimensional vector of a constant size. We feed this vector into a convolutional neural network (CNN) and later into a recurrent neural network (RNN). The CNN in this path is responsible for detecting input-language tokens and patterns, while the RNN is responsible for detecting the sequential order of those patterns. The CNN is composed of eight layers; a combination of one-dimensional convolution, LeakyReLu, and MaxPooling layers. The RNN is a bi-directional Long-Short-Term-Memory (LSTM). This path is highly scalable and requires no prior feature extraction. We achieved good performance using only this path; however, adding the second path provided us with a boost in accuracy performance.

In the second path, count-features are extracted from an input test and fed into a convolution neural network. The sequential order of these count-features is naturally arbitrary; therefore, there is no need for memory elements in the deep neural network (DNN) of this path and a feed forward network is sufficient. We use a convolution neural network to capture any combination of features that might be important for our classification. The CNN in this path is composed of nine layers of a combination of one-dimensional convolution, LeakyReLU, Dense, and MaxPooling Layers. Eventually, both paths are combined using a Softmax layer.

### 3.4   Training and evaluating performance of the DNN

We used the testing techniques to generate millions of samples of the data, which were then classified as benign, malicious, or error. We merged the error and malicious classes. This is a safe practice, since the final results of predicting an error sample as malicious and blocking it, or running it on the application and yielding an error, are essentially the same.

To imitate a real-life situation where vulnerabilities are not yet discovered and evaluate our ahead-of-threat predictions, we split the data into training set and evaluation (testing) set by setting up a time barrier. The time barrier was set such that 99% of the unique test paths fall into the training set, and only 1% fall into the evaluation set. This is an imitation of the real-life continuous testing environment, in which various parties test an application in parallel and the only synchronization point is the Common Vulnerability and Exposure (CVE) database. Next, we used all samples created by the data generation before this time barrier as a training set, and all those created after this time barrier as an evaluation set. We chose not to use random splitting into test and evaluation, since this can create an evaluation set that is very close to the training set, which is not effective in evaluating ahead-of-threat virtual patching.

### 3.5   Real life evaluation

The above evaluation technique is highly dependent on data generation. This is a known limitation of machine learning based on data generation techniques. To overcome this and to evaluate on real life data, we trained the DNN on an old version of the application and used publicly available CVE examples produced years after the release of this version. We then imitated an attacker technique of fuzzing the CVE to get many permutations of the same CVE. We used those permutations to evaluate our ahead-of-threat capability.

To make sure that the ahead-of-threat capability stems up from the predictive DNN and not from the testing algorithm, we executed all the data generated on a new version of the application-under-test and verified they do not expose vulnerability on it.

## 4   Experimental results

To evaluate the system we held the ahead-of-threat patching experiments on two open-source libraries and attempted ahead-of-threat patching on six CVEs. We conducted the experiments on LibXML2 [11] and LibTIFF [10]. **Table 1** shows the version of the library we used for data generation and the version we used for real-life evaluation.

Table 1. Libraries used in the experiment

| Tested library | Version for data generation | Version for evaluation |
|---|---|---|
| LibTIFF | 3.71 | 4.0.7 |
| LibXML2 | 2.6.32 | 2.9.3 |

### 4.1 Data generation

To generate LibXML2 tests we ran AFL on the LibXML2 parser utility testReader from LibXML2 v2.6.32. This testing utility parses input XML files. We ran AFL along with an address sanitizer for three days to generate and classify XML files into three categories: benign, error, and malicious. We used the library's test-corpus as an initial test corpus for AFL and provided XML tokens as a dictionary for AFL. AFL created several millions of non-unique xml files, and the genetic algorithm discovered approximately 200 new tokens. The number of error tests generated far outnumbered the tests in other classes. We merged the error and malicious classes into one class and applied random under-sampling on the error/malicious class. We then calculated the time barrier such that the ratio between unique malicious samples in the training set compared to the evaluation set would be 99 to 1 percent. This process provided us with approximately 600,000 train samples and about 150,000 evaluation samples, where benign and error/malicious classes are equal in the number of samples.

We performed the same process for LibTIFF 3.71 and used the tiffdump utility found in the LibTIFF library. This process provided us with about 300,000 train samples and around 1,100,000 evaluation samples. The large bias towards evaluation samples can be explained by the fact that it is very hard to find new unique tests in later stages of the fuzz testing. Thus, the 99 to 1 percent cut of the unique tests forces an early time barrier on the data generation, while most non-unique tests are found in the later stages.

We limited the size of the numerical vector fed into the first DNN path to 500 and configured the DNN training to run for 4 epochs. We then trained our DNN on the LibXML2 data set using the automatically extracted features and received the following results on the evaluation set. **Table 2** shows the confusion matrix of this evaluation. This yields an accuracy of 89.7% and F1 score of 89.6%.

Table 2. Confusion matrix for LibXML2

|  | Predicted benign | Predicted malicious/error |
|---|---|---|
| True benign | 70,400 | 3,400 |
| True malicious/error | 11,818 | 61,982 |

We performed the same training on the LibTIFF data, and it produced the results shown in **Table 3**. This yields an accuracy of 86.6% and F1 score of 86.6%. In **Fig. 2** we show the receiver operator characteristic (ROC) curves for both models, giving an area under curve (AUC) of 95% for LibXML2 and 98% for LibTIFF.

**Table 3.** Confusion matrix for LibTIFF

|  | Predicted benign | Predicted malicious/error |
|---|---|---|
| True benign | 523,762 | 11,228 |
| True malicious/error | 129,424 | 405,566 |

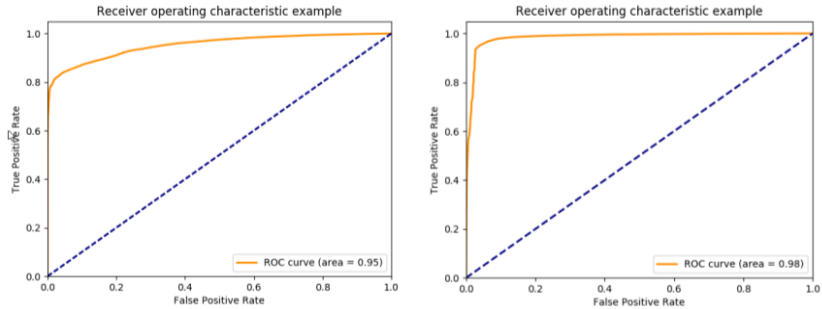

**Fig. 2.** ROC curve for LibXML2 (left, AUC=95%) and LibTIFF (right, AUC = 98%)

### 4.2 Real-life evaluation

We downloaded LibXML2 and LibTIFF exploit proof-of-concept (PoC) files from the publicly available Exploit Database [1]. **Table 4** shows the CVE PoC that we downloaded, date of disclosure, last vulnerable version, release date of the library used for data generation, and the ahead-of-threat years calculated by the difference between the disclosure date and the release date of the library version used for data generation.

**Table 4.** List of vulnerabilities tested with ahead-of-threat patching

| Library | CVE ID | Disclosure date | Last vulnerable version | Release of the library used for data generation | Ahead-of-threat in years |
|---|---|---|---|---|---|
|  | CVE-2015-8806 | 2015-05-08 | 2.9.3 | April 2008 | 7 |
| LibXML2 | CVE-2016-1833 | 2015-11-24 | 2.9.3 |  | 7 |
|  | CVE-2016-1838 | 2016-02-24 | 2.9.3 |  | 8 |
|  | CVE-2017-9147 | 2017-05-12 | 4.0.7 | Dec 2004 | 13 |
| LibTIFF | CVE-2017-9936 | 2017-06-26 | 4.0.8 |  | 13 |
|  | CVE-2017-10688 | 2017-06-29 | 4.0.8 |  | 13 |

We started by validating that the tests we generated in our data-generation phase do not expose any vulnerability in the last vulnerable version of each library. This validated that we are truly generating an ahead-of-threat patching, and not merely an automatic virtual-patch.

Next, we downloaded benign XML and TIFF files from Wikipedia's database XML source code and the FileFormat database [6], respectively. We started to create a test set for the real-life examples by fuzzing the latest vulnerable version of each library using AFL, and the downloaded benign and the CVE PoC files as initial test corpus. This ensured that samples in the malicious class are new tests to our DNN, never found before during the data generation phase, and that the benign and error classes were created from a previously unseen test corpus.

We ran AFL on libXML2 using these initial test corpuses and created tests to get ~20,000 benign tests, ~33,000 error, and ~107,000 malicious tests. Next, we merged the error and malicious classes and under sampled them to get ~20,000 tests. We ran AFL on LibTIFF and got ~10,000 benign, 0 error, and ~18,000 malicious. Since our main attention is on the malicious, we ignored the empty error class and under sampled the malicious to get ~10,000 error/malicious class.

Next, we tested all those samples on the predictive DNN that we trained earlier and received the following results. **Table 5** shows the confusion matrix for LibXML2 results, showing an accuracy of 91.3% and F1 score of 91.3%. **Table 6** shows the confusion matrix for LibTIFF results, showing accuracy of 93.7% and F1 score of 94.3% percent. In **Fig. 3** we show the ROC curves with AUC of 83% for LibXML2 and 96% for LibTIFF. We see that the ROC for LibTiff looks rather smooth, while the ROC for LibXML2 shows a drastic jump around False Positive Rate ~=17%. We attribute this to the difficulty in creating a good real-life evaluation set for a library starting from a small number of CVE examples.

**Table 5.** Real-life evaluation confusion matrix for LibXML2

|  | Predicted benign | Predicted malicious/error |
| --- | --- | --- |
| True benign | 15,622 | 3,192 |
| True malicious/error | 1 | 18,297 |

**Table 6.** Real-life evaluation confusion matrix for LibTIFF

|  | Predicted benign | Predicted malicious/error |
| --- | --- | --- |
| True benign | 105,550 | 6,498 |
| True malicious/error | 9,308 | 131,729 |

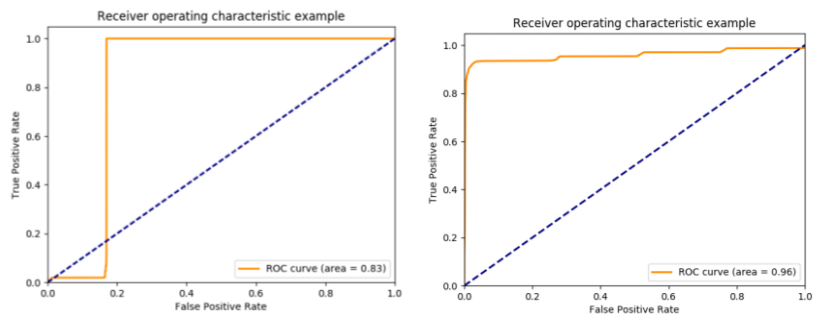

**Fig. 3.** Real life evaluation ROC curve for LibXML2 (left, AUC=83%) and LibTIFF (right, AUC=96%)

## 5     Discussion, conclusion, and future work

We demonstrated how our approach can be used to automatically analyze an application and generate a virtual ahead-of-threat patch. The effort required to test, debug, and fix an application is often very expensive in late stages of the application life cycle. This prevents developers from performing thorough root cause analysis of the application vulnerability. Therefore, they end up adding some error condition that blocks the path to the vulnerability rather than to the vulnerability family. We argue that or approach performs better than the human developer in root cause analysis and in generalizing the benign and error/malicious examples it views to patch a wide family of vulnerabilities. Thus, it does a better job of predicting ahead-of-threat vulnerabilities.

Our next challenge is to apply this approach to virtually patch IoT devices. The need for virtual patches in IoT is even higher in critical devices such as pacemakers. Furthermore, we will study the effect of deploying a virtual patch on a gateway that may block inputs before communicating with the device, compared to deploying a virtual patch on the device itself. Another research direction is expanding the possible targets for patching: our experiments show the effectiveness of this method for C/C++ code; we plan to develop fuzzing techniques for JavaScript programs that will build a system to virtually patch cloud services. Fuzzing of dynamic languages introduces new challenges, such as understanding the underlying memory model. However, we believe it is essential to make our approach applicable to modern software as well.

## 6     Acknowledgements

This project has received funding from the European Union's Horizon 2020 research and innovation programme under grant agreement No 740787 (SMESEC). We would like to thank Ayman Jarrous and Tamer Salman for fruitful discussions, and Ben Liderman for help in building the automated framework.